**Ferromagnetic transition metal implanted ZnO: a diluted magnetic semiconductor?**

Shengqiang Zhou[1,*], K. Potzger[1], Qingyu Xu[1,2], G. Talut[1], M. Lorenz[3], W. Skorupa[1], M. Helm[1], J. Fassbender[1], M. Grundmann[3], H. Schmidt[1]

*1) Institut für Ionenstrahlphysik und Materialforschung, Forschungszentrum Dresden-Rossendorf, Bautzner Landstraße 128, 01328 Dresden, Germany*

*2) Department of Physics, Southeastern University, Nanjing 211189, China*

*3) Institut für Experimentelle Physik II, Universität Leipzig, Linnéstraße 5, 04103 Leipzig, Germany*

Abstract

Recently theoretical works predict that some semiconductors (e.g. ZnO) doped with magnetic ions are diluted magnetic semiconductors (DMS). In DMS magnetic ions substitute cation sites of the host semiconductor and are coupled by free carriers resulting in ferromagnetism. One of the main obstacles in creating DMS materials is the formation of secondary phases because of the solid-solubility limit of magnetic ions in semiconductor host. In our study transition metal ions were implanted into ZnO single crystals with the peak concentrations of 0.5-10 at.%. We established a correlation between structural and magnetic properties. By synchrotron radiation X-ray diffraction (XRD) secondary phases (Fe, Ni, Co and ferrite nanocrystals) were observed and have been identified as the source for ferromagnetism. Due to their different crystallographic orientation with respect to the host crystal these nanocrystals in some cases are very difficult to be detected by a simple Bragg-Brentano scan. This results in the pitfall of using XRD to exclude secondary phase formation in DMS materials. For comparison, the solubility of Co diluted in ZnO films ranges between 10 and 40 at.% using different growth conditions pulsed laser deposition. Such diluted, Co-doped ZnO films show paramagnetic behaviour. However, only the magnetoresistance of Co-doped ZnO films reveals possible s-d exchange interaction as compared to Co-implanted ZnO single crystals.

---

[*] Corresponding author, Email: s.zhou@fzd.de

*Key words:* Diluted magnetic semiconductors, ZnO, Ion implantation, Nanomagnetism

**1 Introduction**

Diluted magnetic semiconductors (DMS) are materials that simultaneously exhibit ferromagnetic and semiconducting properties [1, 2]. They are usually common semiconductor materials containing a few atomic percent of transition metal (TM) ions substituted onto the cation sites. The ferromagnetism in DMS is driven by free charge carriers, and can be controlled by an electrical field. DMS materials could fundamentally change the functionality of traditional semiconductor devices, therefore, have been intensively investigated over the last decades. Among ferromagnetic semiconductors, the (Ga,Mn)As DMS is the most well understood and promising for application in spintronics. The main obstacle is that the highest Curie temperature of (Ga,Mn)As is reported to be 173 K [3], far below room temperature. Nevertheless, spin-related devices based on (Ga,Mn)As, namely spin-polarized light emitter (spin-LED) [4], spin FET [5] and spin-valve [6], have been demonstrated at ***low temperature***. In order to increase the Curie temperature of (Ga,Mn)As, one has to increase the concentration of Mn. However, phase separation, *i.e.* ferromagnetic MnAs precipitates, easily occurs when the Mn concentration is larger than 7% [7]. In parallel, considerable effort is dedicated to search alternative materials, which are expected to be DMS with Curie temperature well above room temperature.

Dietl *et al.* [8] proposed a mean-field Zener model to understand the ferromagnetism in DMS materials. It has been successful in (Ga,Mn)As and (Zn,Mn)Te materials. This model predicts that wide bandgap semiconductors doped with Mn exhibit critical temperatures above 300 K, if a sufficiently large hole density can be achieved ($10^{20}$ cm$^{-3}$). Sato *et al.* calculated the properties of n-type ZnO doped with 3*d* TM ions (V, Cr, Mn, Fe, Co, and Ni) [9]. The ferromagnetic state, with a $T_C$ of around 2000 K, is predicted to be favourable for V, Cr, Fe, Co, and Ni in ZnO while Mn-doped ZnO is predicted to be antiferromagnetic. These predictions largely boosted intensive experimental activities on transition metal doped ZnO. A large number of research groups have reported the experimental observation of ferromagnetism in TM (from Sc to Ni) doped ZnO [10-15] fabricated by various methods including ion implantation. For a comprehensive review, see Ref. [16]. However in these reports the magnetic properties using the same magnetic dopant vary considerably. *E.g.* the saturation moment and Curie temperature for Mn doped ZnO ranges from 0.075 $\mu_B$/Mn, 400 K [11] to 0.17 $\mu_B$/Mn, 30 - 45 K [17], respectively. In contrast to these publications, other groups reported the observation of antiferromagnetism [18-20], spin-glass behavior [21, 22], and paramagnetism [19, 23-25] in TM-doped ZnO. Recently it was also found that nanoscale precipitates can contribute to the ferromagnetic properties substantially. In table 1, we listed the ferro(ferri)-magnetic nanoclusters found in transition metal doped ZnO materials.

The controversy in the magnetic properties of ZnO-based DMS, as stated above, might partially be due to the insufficient characterization of the samples [35-37]. Particularly, a careful correlation between structure and magnetism should be established by sophisticated methods. Synchrotron radiation based x-ray diffraction (SR-XRD) is a powerful tool to detect small precipitates, *e.g.* metallic TM nanocrystals (NC) in ZnO [26]. On the other hand, element selective measurements of the magnetic properties, *e.g.* X-Ray Magnetic

Circular Dichroism (XMCD) [38], and Mössbauer spectroscopy [26, 39], address the origin of the measured magnetism directly.

Among the methods used in the study of ZnO based DMS materials, ion implantation is a non-equilibrium doping method and can overcome the solid-solubility limit of the dopant in substrates. The major drawback of ion implantation is the generation of structural defects in the host lattice. However, in several studies it has been demonstrated that ZnO exhibits a high amorphization threshold. Therefore ion implantation is widely used to dope ZnO with transition metal ions. Ref. [40] gives a review on transition metal ion implantation into ZnO. In this paper, we review our activities on ion implanted ZnO, and scrutinize the formation of secondary phases by a careful correlation between structure and ferromagnetism. Moreover, we compare the preparation method of ion implantation with another non-equilibrium one, pulsed laser deposition (PLD).

## 2 Experimental methods

Commercial ZnO bulk crystals were implanted with $^{57}$Fe, Co, and Ni ions at an elevated temperature of 623 K with fluences from $0.4 \times 10^{16}$ to $8 \times 10^{16}$ cm$^{-2}$. The implantation energy was 180 keV, which results in a projected range of $R_p \sim 89 \pm 29$ nm, and a maximum atomic concentration from 0.5% to 10% (TRIM code [41]).

Structural analysis was achieved both by synchrotron radiation XRD (SR-XRD) and conventional XRD. SR-XRD was performed at the Rossendorf beamline (BM20) at the ESRF with an x-ray wavelength of 0.154 nm. Conventional XRD was performed with a Siemens D5005 equipped with a Cu-target source. In XRD measurements, we use 2θ-θ scans to identify crystalline precipitates, and azimuthal φ-scans for determining their crystallographical orientation.

The lattice damage induced by implantation was evaluated by Rutherford backscattering/channeling spectrometry (RBS/C). $\chi_{min}$ is the channeling minimum yield in RBS/C, which is the ratio of the backscattering yield at channeling condition to that for a random beam incidence [42]. Therefore, $\chi_{min}$ indicates the lattice disordering degree upon implantation.

The magnetic properties were measured with a superconducting quantum interference device (SQUID, Quantum Design MPMS) magnetometer in the temperature range of 5-350 K. Field dependent magnetization (hysteresis loops) was measured at 5 K and 300 K. Temperature dependent magnetization was measured after zero field cooling and field cooling (ZFC/FC) [31]. Note that SQUID magnetometry is an integral method, which measures the total magnetic response from the sample, including substrate and possible contaminations from previous processing [43, 44]. On the other hand, conversion electron Mössbauer spectroscopy (CEMS) is an element specific method, which was used to investigate the $^{57}$Fe lattice sites, electronic configuration and corresponding magnetic hyperfine fields.

The Co-doped ZnO films were grown by pulsed laser deposition (PLD) using a KrF excimer laser. Different substrate temperatures, oxygen pressure and film thickness were chosen to control the electron concentration in the intrinsically n-type conducting ZnO by several orders of magnitude [45].

## 2.1 Fe implanted ZnO

We pick out the ZnO single crystals implanted with Fe as an example to show the possible misinterpretation of the observed ferromagnetism. Fig. 1(a) shows the magnetization measurement on the sample implanted with Fe, with the field along the sample surface. The implantation temperature is 623 K and the Fe fluence is $4\times10^{16}$ cm$^{-2}$. At 5 K and 300 K, the sample shows ferromagnetism. However with increasing temperature, its coercivity and remanence are decreased drastically: from 360 Oe at 5 K to 10 Oe at 300 K, and 0.14 $\mu_B$/Fe to 0.01 $\mu_B$/Fe, and this is a strong indication of superparamagnetism, which has been confirmed by the measurement of ZFC/FC magnetization. The inset of Fig. 1(a) shows ZFC/FC curves with an applied field of 50 Oe. A distinct difference in ZFC/FC curves was observed. ZFC curves show a gradual increase (deblocking) at low temperatures, and reach a broad peak with a maximum, while FC curves continue to increase with decreasing temperature. The broad peak in the ZFC curves is due to the size distribution of Fe NCs.

Fig. 2(b) shows symmetric 2θ-θ scans for the sample performed by conventional XRD and by SR-XRD. Obviously no secondary phases could be detected by conventional XRD, where the sharp peaks, at 2θ ~ 34.4° and 2θ ~ 72.6°, are from bulk ZnO. In contrast to conventional XRD, SR-XRD has the advantages of larger source intensity and signal-background ratio. In the SR-XRD 2θ-θ scan, a rather broad and low intensity peak at 2θ ~ 44.5° originating from α-Fe(110) with a theoretical Bragg angle of 2θ = 44.66° occurs. Apart from α-Fe, no other Fe-oxide ($Fe_2O_3$, $Fe_3O_4$, and $ZnFe_2O_4$) NC are detected. For comparison, the peak intensities for both scans are normalized at the same level. It is clear that the signal-background ratio in SR-XRD is larger than that in conventional XRD by three orders of magnitude. By combining the magnetic and structural measurements, it is reasonable to conclude that metallic α-Fe NC have formed during implantation at 623 K with the fluence of $4\times10^{16}$ $cm^{-2}$, and they are responsible for the ferromagnetism. The saturation moment at 5 K is around 0.24 $\mu_B$/Fe. By comparing with the bulk Fe with a saturation magnetization of around 2.2 $\mu_B$/Fe, around 11% of Fe in this sample is in metallic state. This is further confirmed by CEMS measurements.

Fig. 2 shows the CEMS spectrum for Fe implanted ZnO single crystals at 623 K with a fluence of $4\times10^{16}$ $cm^{-2}$. The majority of Fe are ionic states $Fe^{3+}$ and $Fe^{2+}$, while a considerable fraction of a sextet associated to α-Fe is present. The amount of metallic Fe obtained from CEMS simulation is 12.5%. It is quite agreeable with the results by magnetization measurement. On the other hand, these $Fe^{2+}$ and $Fe^{3+}$ ions could be dispersed inside the ZnO matrix.

Now we would like to find out if lowering of implantation temperatures can avoid the formation of metallic secondary phases. First of all, we check if ion implantation at low temperature results in amorphous ZnO. Figure 3(a) shows the channeling spectra for Fe implanted ZnO at different implantation temperatures. Although the surface damage peak increases drastically with decreasing implantation temperature, the bulk damage peak is hardly affected by implantation temperature. This can be observed clearly in Figure 3(b). The point defects induced by ion-beam can be significantly suppressed by increasing the implantation temperature above 623 K. This temperature is very critical, and below 623 K, the surface damage peak also has no dependence on the substrate temperature. This is very important for the electrical doping of ZnO by ion implantation, where point defects are believed to decrease the conductivity [47].

SR-XRD was performed to check the formation of metallic Fe in the samples implanted at different temperatures from 253 K to 623 K with an Fe fluence of $4\times10^{16}$ $cm^{-2}$. As shown in Figure 4(a), below an implantation temperature of 473 K, no crystalline Fe could be detected. In the 623 K implantation, metallic Fe NCs start to form, therefore the substitution is reduced.

Fig. 4(b) shows the magnetization versus field reversal of samples implanted with Fe ($4\times10^{16}$ $cm^{-2}$) at different implantation temperatures. Only the sample implanted at 623 K shows a hysteretic behavior due to the presence of Fe NCs, while the other samples implanted at 473 K or below show no ferromagnetic response down to 5 K. This is in full agreement with SR-XRD.

With post-implantation annealing, one expects that the metallic Fe NC grow driven by Ostwald rippening. According to magnetization measurements, we found that the annealing at 823 K results in the growth of α-Fe nanoparticles. During annealing at 1073 K the majority of the metallic Fe is oxidized; after a long term annealing at 1073 K, crystallographically oriented $ZnFe_2O_4$ NCs form, which has been reported in Ref. [28].

**2.2 Co and Ni implanted ZnO**

In this section, the formation of Co and Ni NC inside ZnO upon implantation will be discussed. Co or Ni ions were implanted into ZnO at 623 K with the fluence from $0.8\times10^{16}$ to $8\times10^{16}$ cm$^{-2}$. The maximum atomic concentration thus ranges from ~1% to ~10%.

Both conventional XRD and SR-XRD techniques were employed to check the formation of secondary phases in Co or Ni implanted ZnO. Obviously conventional XRD already can detect the formation of metallic Ni NC as shown in Fig. 5(a). At a low fluence ($0.8\times10^{16}$ cm$^{-2}$), no crystalline Ni NC could be detected. At large fluences starting from $4\times10^{16}$ cm$^{-2}$ the Ni(111) peak appears. SR-XRD reveals the same fluence dependence of Ni NC [Fig. 5(b)]. XRD reveals similar results for Co implanted ZnO (not shown), *i.e.* Co NCs start to form at the Co fluence of $4\times10^{16}$ cm$^{-2}$.

Fig. 6 (a), (b) and (c) show the φ-scans of fcc-Ni(200), hcp-Co(10$\underline{1}$1) and ZnO(10$\underline{1}$1), respectively. The azimuthal coordinate (φ) is the angle of rotation about the surface normal. The three φ-scans all exhibit a sixfold symmetry at the same azimuthal position (note that the measurements were only performed with φ ranging of 180°). Therefore, we can conclude that these Co and Ni NCs are crystallographically oriented with respect to the ZnO matrix. The in-plane orientation relationship is hcp-Co[10$\underline{1}$0]∥ZnO[10$\underline{1}$0], and Ni[112]∥ZnO[10$\underline{1}$0], respectively. Due to the hexagonal structure of Co and three-fold symmetry of Ni viewed along [111] direction, it is not difficult to understand their crystallographical orientation onto hexagonal-ZnO.

Correspondingly, magnetization measurements reveal similar fluence dependence of the formation of metallic Co or Ni NC. Superparamagnetism was measured for the samples with fluences of $4\times10^{16}$ and $8\times10^{16}$ cm$^{-2}$, not for the fluence of $0.8\times10^{16}$ cm$^{-2}$. The detailed structural and magnetic properties as well as annealing behavior can be found in ref. [30].

## 2.3 PLD grown ZnCoO films

For comparison, we have also investigated Co doped ZnO thin films grown by PLD. The crystal structure of the films was characterized by x-ray diffraction measurements with 2θ-θ scans using a Cu K$_\alpha$ source. As a sharp contrast with Co implanted ZnO, no metallic Co clusters are formed in ZnCoO films. Only (0002) and (0004) peaks of wurtzite ZnO were observed, indicating the highly c-axis-oriented magnetic ZnO films without any visible impurities [48]. Note that Co NC are crystallographically oriented inside ZnO matrix, and they are rather easy to be detected. The structure of the films was also studied by transmission electron microscopy. No impurities were observed. The Co$^{2+}$ distribution was uniform in the film as observed by electron energy loss microscopy EELS and elemental mapping [49]. Therefore it is reasonable to assume that Co ions are diluted in PLD grown ZnO films.

According to the theory of Dietl *et al.*[8], p-d interactions are the reason for long-range magnetic coupling. However, the investigated magnetic ZnO samples are either n-type conducting or insulating. The observed weak ferromagnetism in implanted ZnO containing ferromagnetic NC and in PLD grown ZnO with diluted magnetic ions is due to ferromagnetic NC and the acceptor-like defects [50, 51], respectively. For the diluted ZnCoO, it would be interesting to check if there is interaction between free electrons and d-electrons in Co ions, and if this s-d interaction results in ferromagnetic coupling. In doped ZnO, the charge transport is strongly affected by quantum interference of both scattered waves and amplitudes of the electron-electron interaction, which in turn results in different magnetoresistance (MR) effects [52]. In Ref. [53], we have determined the critical electron concentration at the metal-insulator transition, $n_c$ of $4\times10^{19}$ cm$^{-3}$. Below $n_c$, the character of wave functions changes from delocalized to localized. We studied the MR effect in Co-doped ZnO films with electron concentration ranging from $8.3\times10^{17}$ to $9.9\times10^{19}$ cm$^{-3}$ (around the metal-insulator transition) experimentally and theoretically. We attempt to find if there is an indication of s-d interaction in the Co doped ZnO films.

Figure 7 shows the MR measurement in Co-doped ZnO with different electron concentrations [53]. A large positive MR of 124% has been observed in the film with the lowest electron concentration of $8.3\times10^{17}$ cm$^{-3}$, while only negative MR of -1.9% was observed in the film with an electron concentration of $9.9\times10^{19}$ cm$^{-3}$ at 5 K [54]. The positive MR decreases drastically with increasing temperature, and only negligible negative MR can be observed above 50 K. In magnetic doped ZnO, the positive MR is related with the quantum corrections to the conductivity due to the influence of the spin splitting of the conduction band on the electron-electron interaction [52, 55]. For negative MR the underlying physical origin is much debated. We previously observed negative MR in Ti-, Cu-, and Nd-doped ZnO, and modelled it by considering the magnetic scattering of the conduction electrons by isolated magnetic ions, as proposed by M. Csontos et al. [56]. We applied the Csontos model to Co-doped ZnO and observed that above 50 K the negative MR in Co-doped ZnO is not mainly influenced by magnetic impurity scattering. For semiconductors in the weak localization region, the field suppression of weak localization is a possible origin for the small negative MR [57], which has been applied to explain the small negative MR in Co-doped ZnO. We combined the quantum correction of s-d spin-splitting on the disorder modified electron-electron interaction and the field suppression of weak localization to fit the MR in Co-doped ZnO [54]. The solid lines in Fig. 7 are the fitting results. The agreement between modelled and experimental MR data is excellent. In ref. [52], the presence of a giant spin-splitting is specific to DMS in a paramagnetic phase. Actually we do only observe paramagnetism in Co doped ZnO films. Fig. 8(a) shows a typical M-H curve for the Co-doped ZnO sample measured at 2 K. The magnetization slowly gets saturated at high field, and neither coercivity nor remanence can be observed. Paramagnetism has been well described by Brillouin function. The substituted $Co^{2+}$ ions might have two states, one with $L=3$ and $S=3/2$, and the other with $L=1.07$, $S=3/2$ [58]. It can be clearly seen that with $L=1.07$ and $S=3/2$ the Brillouin function can fit the M-H curve at 2 K well. Correspondingly the ZFC/FC curves [Fig. 8(b)] are completely overlapped with each other, confirming that there is no ferromagnetism in this sample. The detailed results will be published elsewhere [59].

## 3 Conclusions

Concluding the presented results, we have shown that ferromagnetic secondary phases, namely metallic Fe, Co and Ni, as well as Zn-ferrites, have formed upon ion implantation into ZnO bulk crystals and post annealing. These secondary phases are responsible for the observed ferromagnetism. We did not observe any indication for ferromagnetic DMS created by our methods.

By PLD, we have prepared a single phase of $Zn_{1-x}Co_xO$ with x up to 0.1. However only paramagnetism was observed, while magnetoresistance reveals the possible s-d exchange interaction.

Other groups working with different preparation methods experience similar behaviors, where diluted $Zn_{1-x}Co_xO$ films can be prepared with excellent crystalline quality, but only paramagnetism has been observed [58, 60].

## 4 Acknowledgement

This work is partially (S. Z., Q. X., and H. S.) funded by BMBF (Grant No. FKZ03N8708). Q. X. acknowledges the National Natural Science Foundation of China (50802041).

Fig captions

Figure 1 (color online): (a)Magnetization reversal recorded at 5 and 300 K using SQUID magnetometry for the sample implanted with Fe at 623 K and the fluence of $4\times10^{16}$ cm$^{-2}$. Inset shows the ZFC/FC curves with an applied field of 50 Oe for the same sample. (b) Conventional (Conv.) and SR-XRD pattern (2θ-θ scan) for the Fe implanted ZnO. Adapted from Ref. [46].

Figure 2 (color online): Room temperature CEMS of ZnO bulk crystals implanted with $^{57}$Fe at a temperature of 623 K. Adapted from Ref. [46].

Figure 3 (color online): (a) Representative RBS/C spectra with different implantation temperature. The fluence is $4\times10^{16}$ cm$^{-2}$, and implantation energy is 180 keV. (b) The calculated $\chi_{min}$ for different implantation temperature, Implantation at low temperature (473 K) results in more damage at the surface region. Adapted from Ref. [46]

Figure 4 (color online): (a) SR-XRD 2θ-θ scans of Fe implanted ZnO with the same fluence of $4\times10^{16}$ cm$^{-2}$ at difference temperature. In order to show the Fe(110) peak, the figure is spliced to two parts. Only the sample implanted at 623 K shows α-Fe precipitates. (b) Hysteresis loops measured at 5 K for Fe implanted ZnO at different implantation temperatures. Only the sample implanted at 623 K shows a hysteresis behavior. Adapted from Ref. [46].

Figure 5 (color online): 2θ-θ scan revealing the formation of metallic Ni NC in Ni implanted ZnO (a) conventional XRD, and (b) SR-XRD. (The fluence for Ni ions is indicated). Adapted from Ref. [30].

Figure 6 (color online): XRD φ-scans revealing the crystallographical orientation relationship between Co/Ni NCs and ZnO matrix: (a) Ni(200); (b) hcp-Co(10$\underline{1}$1) and (c) ZnO(10$\underline{1}$1).

Figure 7: The magnetoresistance vs. magnetic field (open symbols) for the samples with different electron concentrations (a) $9.9\times10^{19}$ cm$^{-3}$, (b) $1.7\times10^{19}$ cm$^{-3}$, (c) $5.1\times10^{18}$ cm$^{-3}$, and (d) $8.3\times10^{17}$ cm$^{-3}$, at different temperatures. The solid lines are the fitting curves. Adapted from Ref. [54].

Figure 8 (color online): (a) The M-H curves measured at 2K for a typical ZnCoO films and the fitting using Brillouin function with $L$=1.07 and $S$=3/2. (b) Temperature dependent magnetization: the ZFC curve is overlapped with the FC curve.

Table 1: Second phases observed in TM-doped ZnO and their magnetic properties. Curie temperature (for ferro- or ferrimagnetic material) of these secondary phases in bulk form is given.

| Secondary phase | Magnetism | Curie temperature | Reference |
|---|---|---|---|
| Fe | Ferromagnetic | 800 K | [26] |
| $ZnFe_2O_4$ (inverted) | Ferrimagnetic | | [27, 28] |
| Co | Ferromagnetic | 1373 K | [29,30] |
| Ni | Ferromagnetic | 630 K | [31] |
| $(Zn,Mn)Mn_2O_4$ | Ferrimagnetic | 40 K | [32] |
| $Mn_3O_4$ | Ferromagnetic | 43 K | [33] |
| $Mn_{2-x}Zn_xO_{3-\delta}$ | Ferromagnetic | 980 K | [33] |
| CoZn | Ferromagnetic | 400-450 K | [34] |

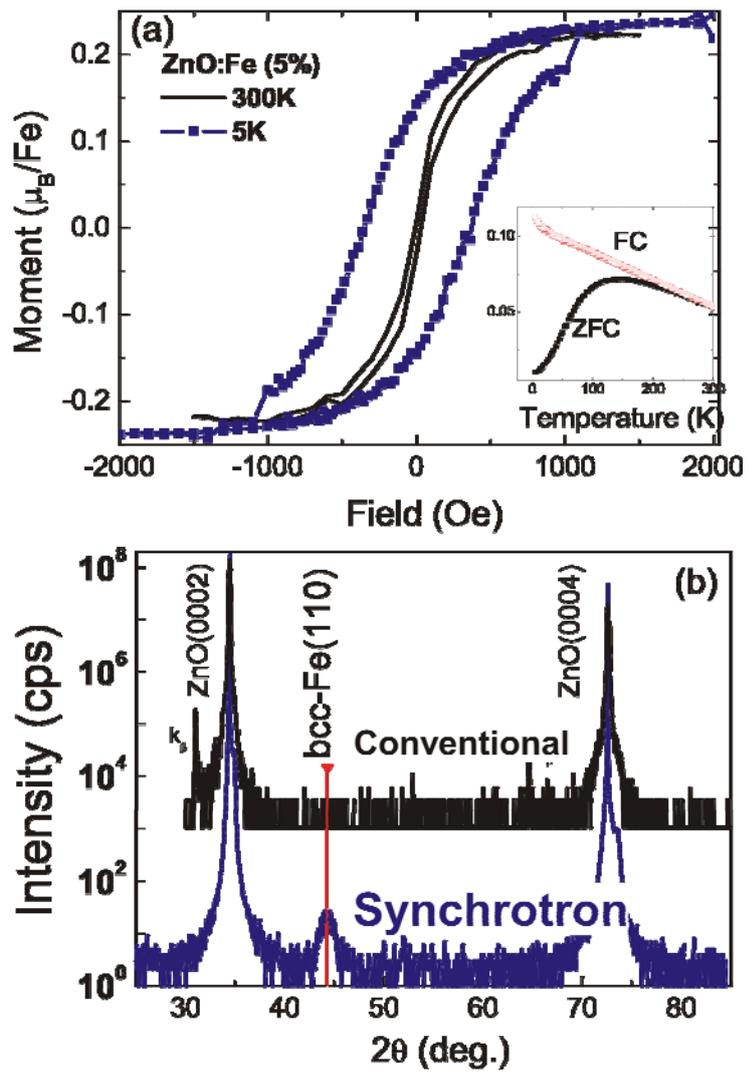

Fig. 1

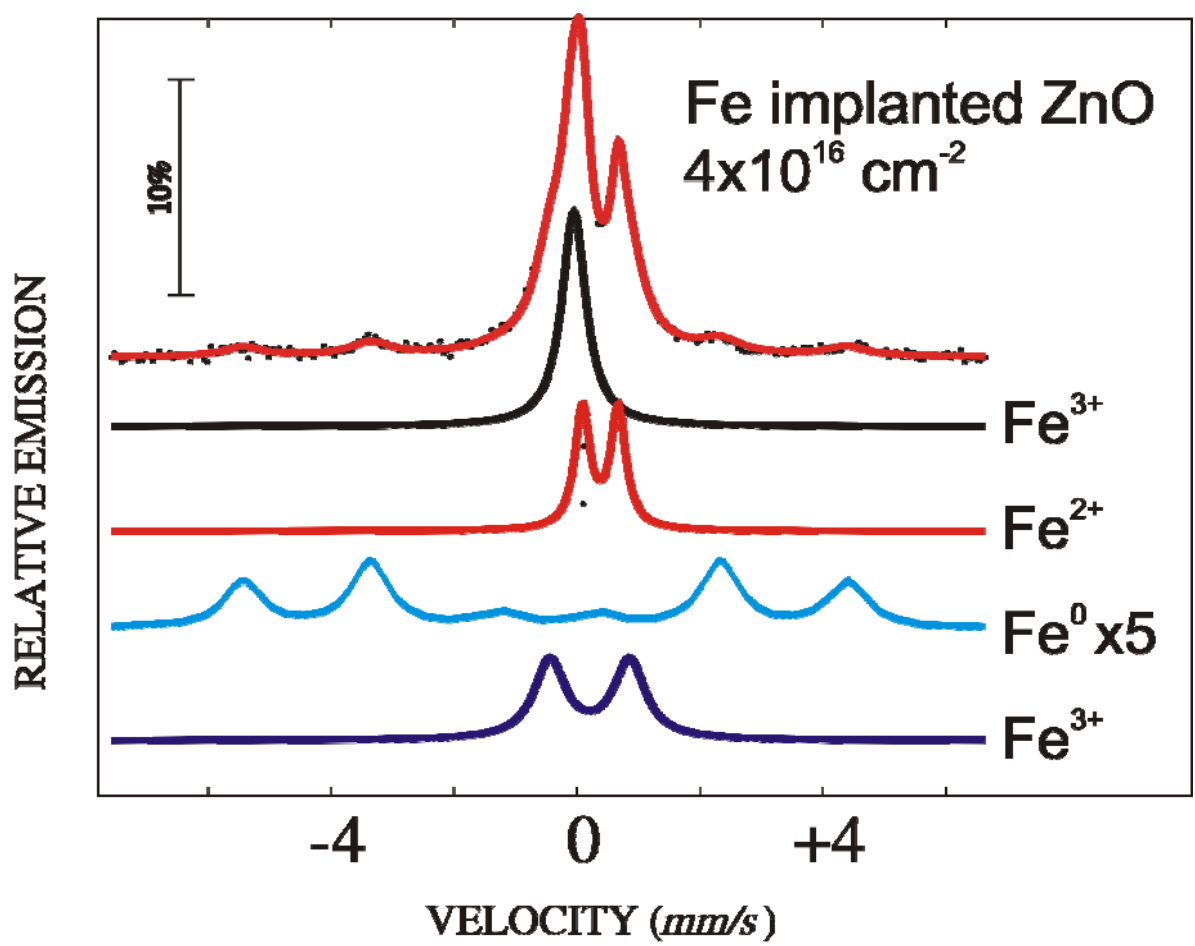

Fig. 2

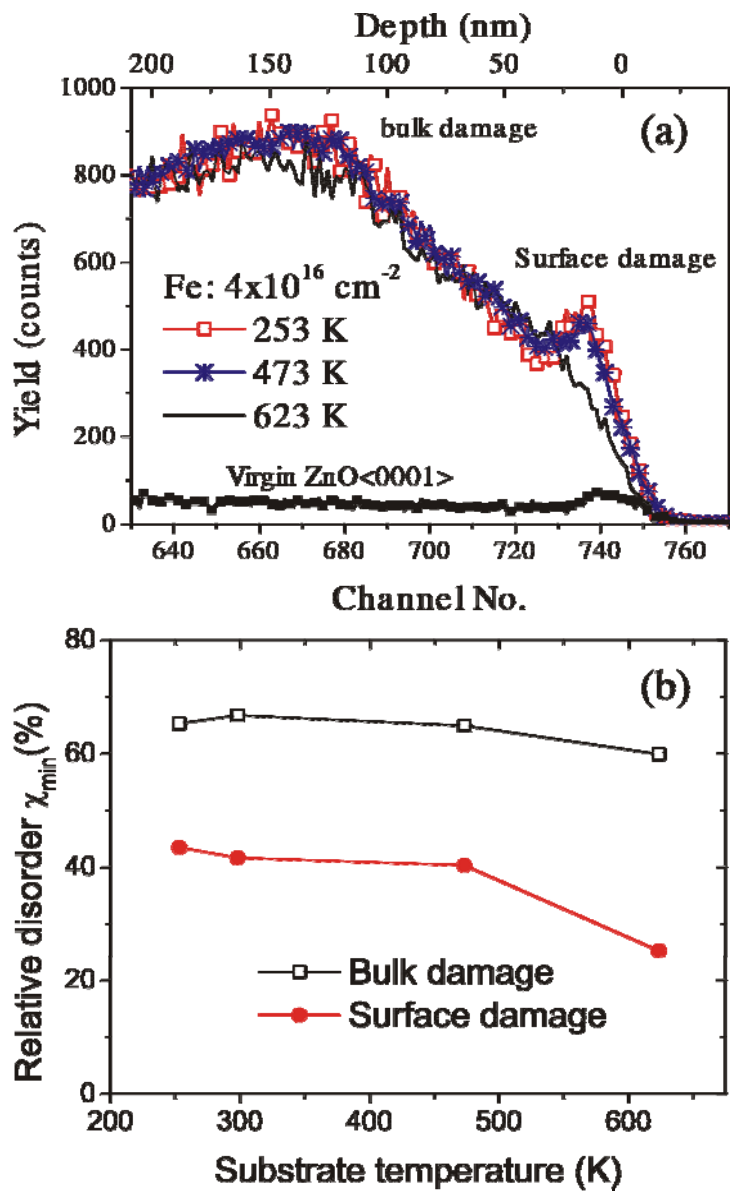

Fig. 3

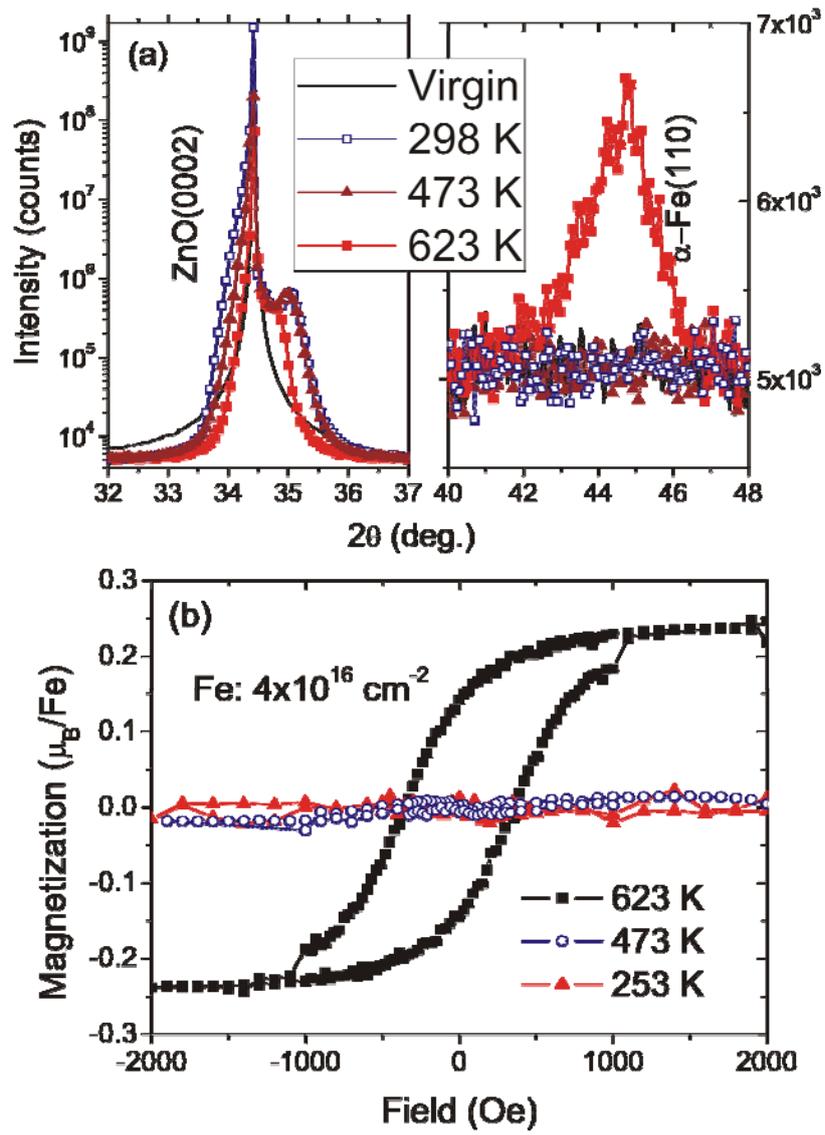

Fig. 4

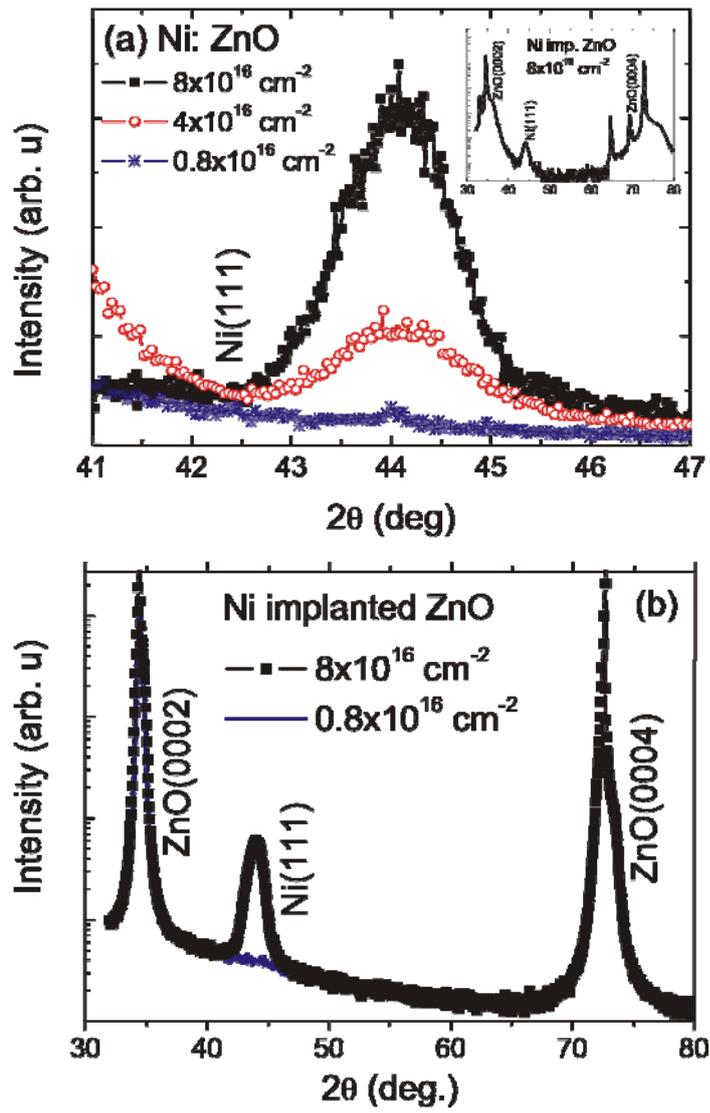

Fig. 5

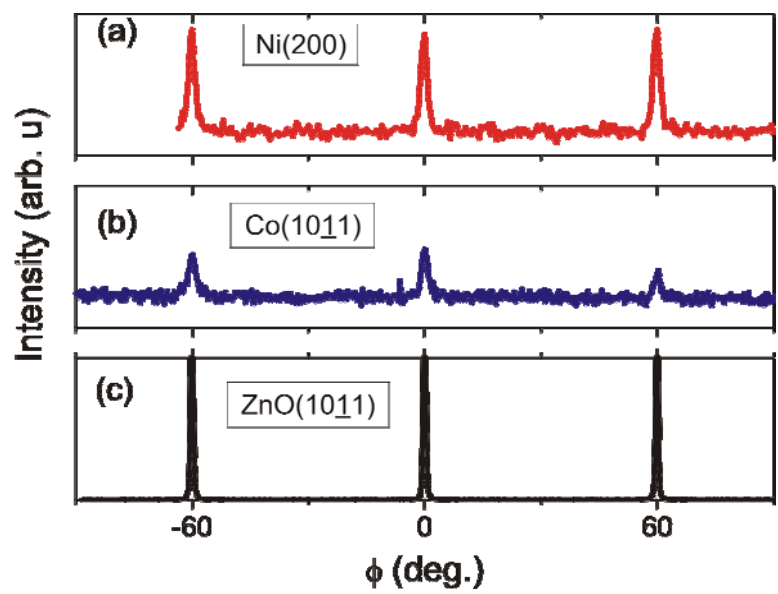

Fig. 6

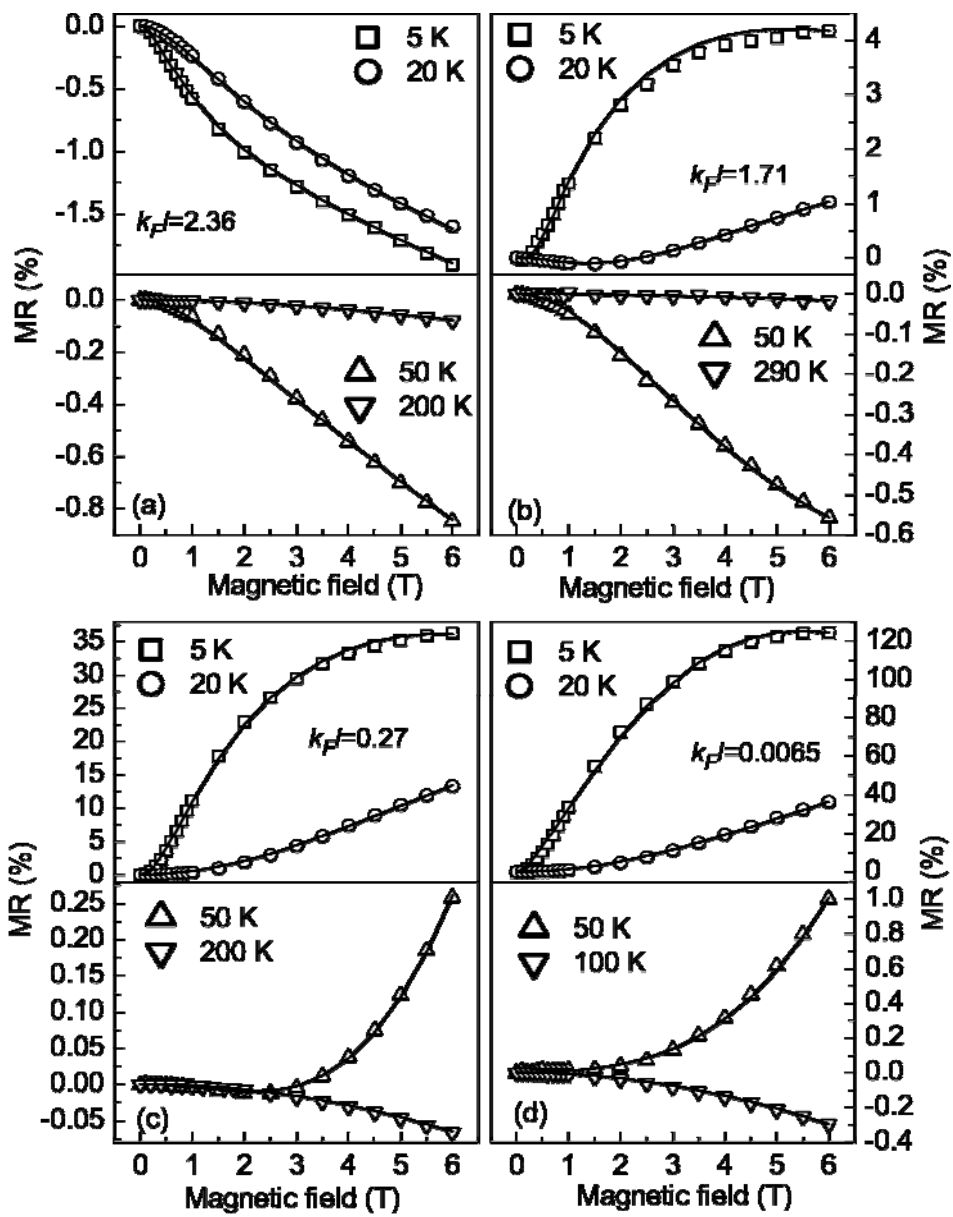

Fig. 7

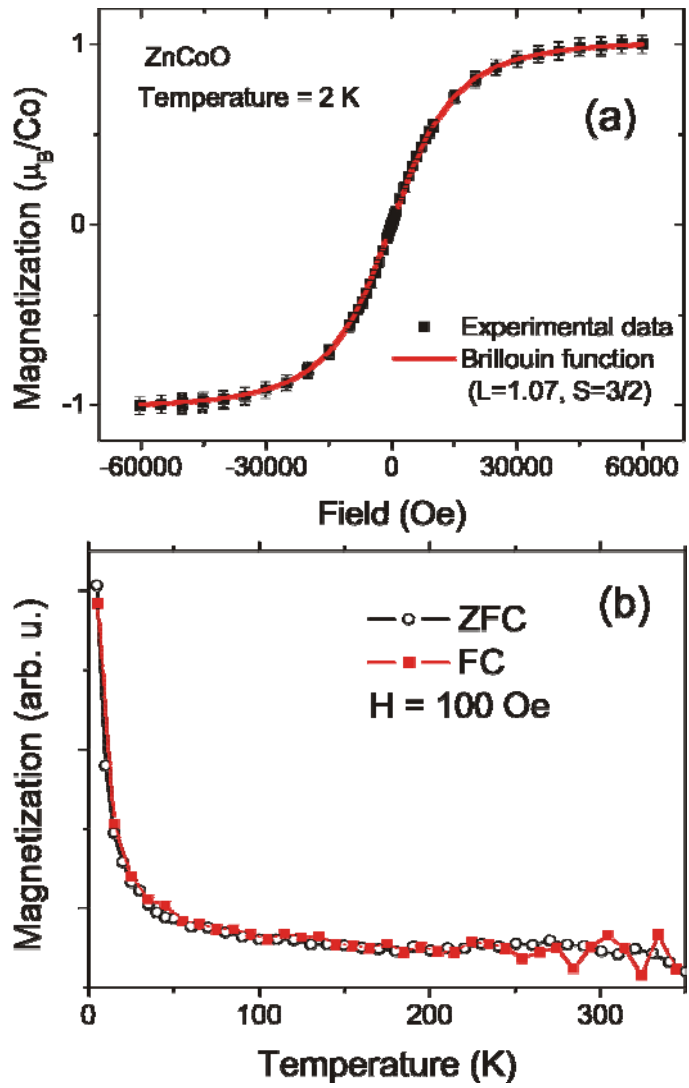

Fig. 8